# On Mixed Quantum/Classical Theory for Rotationally Inelastic Scattering of Identical Collision Partners

D. Bostan, B. Mandal and D. Babikov[1]

**Abstract:** Mixed quantum/classical theory (MQCT) for the treatment of rotationally inelastic transitions during collisions of two identical molecules, described either as indistinguishable or distinguishable partners, is reviewed. The treatment of two molecules as indistinguishable includes symmetrization of rotational wavefunctions, introduces exchange parity, and gives state-to-state transition matrix elements different from those in the straightforward treatment of molecules as distinguishable. Moreover, the treatment of collision partners as indistinguishable is eight times faster. Numerical results, presented for $H_2 + H_2$, $CO + CO$ and $H_2O + H_2O$ systems, indicate good agreement of MQCT calculations with full-quantum calculations from literature and show that an *a posteriori* correction, applied after the treatment of collision partners as distinguishable, generally produces good results that agree well with the rigorous treatment of collision partners as indistinguishable. This correction for cross section includes either multiplication by 2, or summation over physically indistinguishable processes, depending on the transition type. After this correction, the results of two treatments agree within 5% for most but may reach 10-20% for some transitions. At low collision energies dominated by scattering resonances these differences can be larger, but they tend to decrease as collision energy is increased. It is also shown that if the system is artificially forced to follow the same collision path in the indistinguishable and distinguishable treatments, then all differences between the results of two treatments disappear. This interesting finding gives new insight into the collision process and indicates that indistinguishability of identical collision partners comes into play through the collision path itself, rather than through matrix elements of inelastic transitions.

**TOC:** If the system is artificially forced to follow the same collision path in the indistinguishable and distinguishable treatments, then all differences between the results of two treatments disappear.

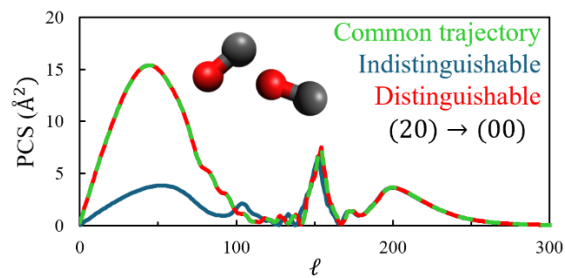

*Chemistry Department, Marquette University, Milwaukee, Wisconsin 53201-1881, USA.*
[1] *E*-mail: dmitri.babikov@mu.edu

## I. INTRODUCTION

Collisions of identical molecules play important roles in various environments in the universe. Since $H_2$ is the most abundant molecule in space, the process of rotational energy transfer in $H_2 + H_2$ collisions is of key importance for the modeling of star forming regions,[1] and therefore it has been extensively studied using both theory[2–6] and experiments.[7–10] The second most abundant molecule in space is CO.[1,11] Describing the process of CO + CO collisions is important for the modelling of cometary comas, in particular, for the comets that originate in the most distant and coldest parts of the Solar system (in the Oort Cloud)[11] and for the interstellar comets that come from the outside of Solar system.[12–14] For the comets of Kuiper Belt (closer to Earth) the collisions between water molecules, $H_2O + H_2O$, are expected to play important role in radiation transfer under non-equilibrium conditions.[11] The modelling of spacecraft entry into Martian atmosphere requires the knowledge of $CO_2 + CO_2$ energy transfer cross sections,[15,16] that are also important for the atmospheres of many exoplanets. And of course, inelastic and dissociative $N_2 + N_2$ and $O_2 + O_2$ collisions represent some of the most important processes in Earth's atmosphere.[17–19] In all these and many other cases one must deal with identical collision partners.

When two identical molecules collide, the wavefunction of the overall system must satisfy certain properties imposed by symmetry. Namely, the probability distribution should remain unchanged if two identical molecules are swapped (exchanged). This property was incorporated into quantum scattering theories by different authors, but in several alternative ways that lead to several different versions of the final equation.[20–22] This issue created some confusion in the literature concerning the inclusion of factor of 2 into the formula for state-to-state transition cross sections.[23,24] A user-ready inelastic scattering code, MOLSCAT,[25,26] has an option of treating two colliding molecules as indistinguishable, still, some confusion remains and the debates about the inclusion of the factor of 2 continue, up to this day.[27,28] But, besides the factor of 2, it also remains unclear how important is the incorporation of collision partners exchange symmetry and what effect it has onto computed state-to-state transition cross sections. Is it always necessary to include the exchange symmetry of identical collision partners in the calculations, or can we neglect it at least in certain cases?

One of the goals of this paper is to carry out the calculations of rotationally inelastic scattering of identical molecules in two different ways: one treating collision partners rigorously as *indistinguishable* with incorporation of exchange symmetry into the wavefunction of the system, and the other without it, treating two collision partners as *distinguishable*. It is interesting to find out how different the results of two approaches are, how this difference depends on collision energy, and how it changes when we go from one molecule to another (here we consider $H_2 + H_2$, $CO + CO$, and $H_2O + H_2O$ systems). For this, we use our recently developed mixed quantum/classical theory,[29,30] which treats rigorously the rotational motion of two molecules using time-dependent Schrodinger equation but employs Ehrenfest approximation to describe the relative translational motion of collision partners (their scattering) using mean-field trajectories. This method is implemented in the code MQCT[31–33] which permits to efficiently compute state-to-state transition cross sections at different collision energies and for different molecular systems. Another important goal of this paper is to ensure that the predictions of MQCT code are consistent with results of full-quantum methods, including the aforementioned factor of 2.

## II. THEORY

In this section we review those elements of the mixed quantum/classical theory of inelastic scattering that are relevant to the property of indistinguishability of collision partners and make difference in the cases of *distinguishable* and *indistinguishable* collision partners. This includes rotational wavefunctions of molecular eigenstates, matrix elements for rotational state-to-state transitions, and the formulae for calculations of state-to-state transition cross sections. MQCT equations of motion remain the same in both cases and are not reviewed here for the purpose of brevity (interested readers are encouraged to consult several recent papers).[34–37] For simplicity, the theory is presented for a linear-rotor + linear-rotor case, appropriate for diatom + diatom systems such as $H_2 + H_2$ and $CO + CO$. The most general case of asymmetric-top rotor + asymmetric-top rotor (such as $H_2O + H_2O$) is mentioned where applicable.

### II-A. Rotational wavefunctions

In MQCT the rotational motion of each molecule is described by active Euler rotations relative to the molecule-molecule vector that connects centers-of-masses of two collision partners and plays the role of the quantization axis (known as body-fixed reference frame).[29] Then, $\Lambda_1 = (\alpha_1, \beta_1, \gamma_1)$ and $\Lambda_2 = (\alpha_2, \beta_2, \gamma_2)$ represent two sets of Euler angles that permit to describe rotations of any two molecules. For linear rotors, only two Euler angles are needed, two quantum numbers are used, and the rotational wavefunctions are known analytically (spherical harmonics): $Y_{m_1}^{j_1}(\beta_1, \alpha_1)$ and $Y_{m_2}^{j_2}(\beta_2, \alpha_2)$ for molecules "1" and "2". Then, the rotational wavefunctions of the overall molecule-molecule system are given by the following expression:[38]

$$|j_1 j_2 j m\rangle = \sum_{m_1=-j_1}^{+j_1} C_{j_1,m_1,j_2,m-m_1}^{j,m} Y_{m_1}^{j_1}(\beta_1, \alpha_1) Y_{m-m_1}^{j_2}(\beta_2, \alpha_2) \qquad (1)$$

Here $C_{j_1,m_1,j_2,m_2}^{j,m}$ are Clebsh-Gordon coupling coefficients, $j$ is the overall angular momentum quantum number (sometimes called $j_{12}$ in the literature) varied through the range $|j_1 - j_2| \leq j \leq j_1 + j_2$, and $m$ is projection of $j$ onto the body-fixed quantization axis varied through the range $-j \leq m \leq +j$. Overall, this gives $(2j_1 + 1)(2j_2 + 1)$ quantum states of the molecule-molecule system for any chosen values of $j_1$ and $j_2$ of the individual molecules, that determine what we will call a "channel" $(j_1 j_2)$. Note that $m_2 = m - m_1$ is used in Eq. (1) since only in these cases the values of Clebsh-Gordon coefficients are different from zero. Wavefunctions of Eq. (1) are used in the case of *distinguishable* collision partners, and will be labelled below as $|j_1 j_2 j m\rangle_{\Lambda_1 \Lambda_2}$, which emphasizes that molecules "1" and "2", described by the sets of Euler angles $\Lambda_1$ and $\Lambda_2$, are placed in quantum states $j_1$ and $j_2$, respectively.

We can also consider an alternative, energetically equivalent set of states given by:[38]

$$|j_1 j_2 j m\rangle_{\tilde{\Lambda}_2 \tilde{\Lambda}_1} = \sum_{m_1=-j_1}^{+j_1} C_{j_1,m_1,j_2,m-m_1}^{j,m} Y_{m_1}^{j_1}(\beta_2, \alpha_2) Y_{m-m_1}^{j_2}(\beta_1, \alpha_1) \qquad (2)$$

where molecules "2" and "1" described by the sets of Euler angles $\Lambda_2$ and $\Lambda_1$, are placed in quantum states $j_1$ and $j_2$, respectively (in this order). Tildes are used to emphasize that this configuration is obtained by swapping molecules "1" and "2". In the case of *indistinguishable* collision partners, the two choices, Eqs. (1) and (2), are equally possible and two sets of symmetrized wavefunctions can be obtained by superpositions:[38]

$$|j_1 j_2 j m\rangle^{\pm} = \frac{|j_1 j_2 j m\rangle_{\Lambda_1 \Lambda_2} \pm |j_1 j_2 j m\rangle_{\tilde{\Lambda}_2 \tilde{\Lambda}_1}}{\sqrt{2(1 + \delta_{j_1 j_2})}} \qquad (3a)$$

Besides the usual quantum numbers, these states are labeled by positive or negative *exchange parity* (±) that corresponds to the sign in the numerator of Eq. (3a). $\delta_{j_1 j_2}$ is the usual Kronecker symbol that turns to one if two quantum states are identical but is zero other vice. It is possible to show that this expression can be rewritten in the following form:[38]

$$|j_1 j_2 j m\rangle^{\pm} = \frac{|j_1 j_2 j m\rangle_{\Lambda_1 \Lambda_2} \pm p |j_2 j_1 j m\rangle_{\Lambda_1 \Lambda_2}}{\sqrt{2(1 + \delta_{j_1 j_2})}} \tag{3b}$$

Note that in this formula the order of angles $\Lambda_1$ and $\Lambda_2$ is the same in both terms, but $j_1$ and $j_2$ are swapped in the second term, which is a shortened notation that means that we use quantum state $j_2$ for molecule 1, whereas quantum state $j_1$ is used for molecule 2. Importantly, the factor $p$ which appears in front of the second term corresponds to the *inversion parity* of the rotational state of the molecule-molecule system as a whole (the overall inversion parity). For a system of two linear-top rotors, $p = (-1)^j$. For a system of two asymmetric-top rotors (as $H_2O + H_2O$) the expression is more complicated:[34] $p = (-1)^j (-1)^{\kappa_1 + \kappa_2} \varepsilon_1 \varepsilon_2$, where $\kappa_1$ and $\kappa_2$ correspond to the para/ortho character (0 or 1) of the two states, while $\varepsilon_1$ and $\varepsilon_2$ correspond to their inversion parities ($\pm 1$).

It should be stressed that in the case when two molecules are in the same state, $j_1 = j_2$ (which we will call a "pair" state, following Ref. [28]) the value of $\pm p$ can only be positive, because the negative value will annihilate the wavefunction. So, in the case of pair states only positive exchange parity (+) is possible with positive inversion parity $p = 1$ (for even $j$ states) and only negative exchange parity (−) is possible with negative inversion parity $p = -1$ (for odd $j$ states within a channel).

It should also be noted that since the definition of the overall inversion parity $p$ includes $(-1)^j$, and since the value of $j$ is not fixed for a given channel $(j_1, j_2)$ but is varied through the range $|j_1 - j_2| \leq j \leq j_1 + j_2$, we will have states with both positive and negative values of $p$ within the manifold of rotational states produced by the choice of $(j_1, j_2)$. Those states with even values of total $j$ will have $p = 1$, but those with odd values of total $j$ will have $p = -1$. The only exception is the trivial situation when, say, $j_1 = 0$, and therefore only one value of $j$ is possible, $j = j_2$, in which case only one value of inversion parity is possible (that depends on whether the value of $j_2$ is even or odd).

### II-B. State-to-state transition matrix elements

In the case of *distinguishable* collision partners the wavefunctions of Eq. (1) are used to compute matrix elements of potential coupling for state-to-state transitions, as follows:

$\langle j'_1 j'_2 j' m | V(R, \Lambda_1, \Lambda_2) | j_1 j_2 j m \rangle$, where prime is used to indicate quantum numbers of the final states, while $V(R, \Lambda_1, \Lambda_2)$ represents the potential energy surface of the system, with $R$ being the molecule-molecule distance. The potential coupling matrix is diagonal in $m$, but the values of matrix elements depend on the order of states of distinguishable collision partners. Namely, in general, the matrix element given above is different from the matrix elements where the (initial or final) states of two molecules are swapped, such as $\langle j'_1 j'_2 j' m | V(R, \Lambda_1, \Lambda_2) | j_2 j_1 j m \rangle$ or $\langle j'_2 j'_1 j' m | V(R, \Lambda_1, \Lambda_2) | j_1 j_2 j m \rangle$.

In the case of *indistinguishable* collision partners the wavefunctions of Eq. (3) are used to compute matrix elements:

$$\langle j'_1 j'_2 j' m | V(R, \Lambda_1, \Lambda_2) | j_1 j_2 j m \rangle^\pm$$

$$= \frac{1}{\sqrt{2(1 + \delta_{j_1 j_2}) 2(1 + \delta_{j'_1 j'_2})}} \begin{bmatrix} \langle j'_1 j'_2 j' m | V(R, \Lambda_1, \Lambda_2) | j_1 j_2 j m \rangle \\ \pm\ p\ \times \langle j'_1 j'_2 j' m | V(R, \Lambda_1, \Lambda_2) | j_2 j_1 j m \rangle \\ \pm\ p' \times \langle j'_2 j'_1 j' m | V(R, \Lambda_1, \Lambda_2) | j_1 j_2 j m \rangle \\ +\ pp' \times \langle j'_2 j'_1 j' m | V(R, \Lambda_1, \Lambda_2) | j_2 j_1 m \rangle \end{bmatrix} \quad (4)$$

Here $p$ and $p'$ represent the inversion parity (defined above) for the initial and final states of a two-molecule system. The four terms correspond to all possible combinations of swaps of the initial and final states and describe four physically indistinguishable transitions: $(j_1 j_2) \to (j'_1 j'_2)$, $(j_2 j_1) \to (j'_1 j'_2)$, $(j_1 j_2) \to (j'_2 j'_1)$ and $(j_2 j_1) \to (j'_2 j'_1)$ but, the symmetry of interaction potential permits to simplify this expression, as follows:

$$\langle j'_1 j'_2 j' m | V(\Lambda_1, \Lambda_2) | j_1 j_2 j m \rangle^\pm$$

$$= \frac{1}{\sqrt{(1 + \delta_{j_1 j_2})(1 + \delta_{j'_1 j'_2})}} \left[ \begin{matrix} \langle j'_1 j'_2 j' m | V(\Lambda_1, \Lambda_2) | j_1 j_2 j m \rangle \\ \pm p \langle j'_1 j'_2 j' m | V(\Lambda_1, \Lambda_2) | j_2 j_1 j m \rangle \end{matrix} \right] \quad (5)$$

Here we adopted the form where a swap in the initial states is used, with one final state. This formula is implemented in the MQCT code. An equivalent expression where only the final state quantum numbers are swapped can also be employed.

It should be noted that only transitions between states within the same exchange parity are allowed (the parity changing transitions are forbidden). Therefore, the calculation of two matrix elements for *indistinguishable* collision partners, those two in Eq. (5) for transitions between positive and negative exchange parity states ($\pm$), requires the same numerical effort as calculations of two matrix elements for transitions $(j_1 j_2) \to (j'_1 j'_2)$ and $(j_2 j_1) \to (j'_1 j'_2)$ in the case of distinguishable collision partners. But, if the two molecules are treated as *distinguishable* (as if they would be entirely different, say CO + $H_2$), we would also have to compute two more matrix elements, those for transitions $(j_1 j_2) \to (j_2' j_1')$ and $(j_2 j_1) \to (j_2' j_1')$ since they represent physically different processes with different potential couplings. This is easy to understand if we pick up a set of arbitrary quantum numbers for, say CO + $H_2$, and try to swap them: for example $(0,2) \to (4,6)$, $(2,0) \to (4,6)$, $(0,2) \to (6,4)$ and $(2,0) \to (6,4)$. One can immediately see that neither of these transitions are equivalent. Therefore, the calculation of matrix elements for *indistinguishable* collision partners is a factor of two cheaper, compared to the case of *distinguishable* molecules. This is true for non-pair ↔ non-pair, and for pair ↔ non-pair transitions. The only case when the numerical effort for computing matrix elements is the same in distinguishable and indistinguishable approaches corresponds to pair ↔ pair transitions such as $(2,2) \to (4,4)$, when only one matrix element is needed in either case, but these transitions are relatively rare compared to the other two types of transitions.

The other source of different numerical cost in distinguishable and indistinguishable treatments comes from a set of states included (and the number of matrix elements used) in the actual calculations of molecule-molecule collision. Namely, if the collision partners are treated as *distinguishable* then all states need to be included in the basis set expansion of rotational wavefunction, and all transitions considered at the same time. In the example of CO + $H_2$ considered above, this would include all four states: $(0,2), (2,0), (4,6)$ and $(6,4)$ with all possible state-to-state transition matrix elements. In the case of *indistinguishable* partners we have to run two separate calculations of molecule-molecule collision: one with only positive exchange parity states in the basis (i.e., symmetrized $(0,2)^+$ and $(4,6)^+$ with their corresponding transition matrix element) and the other with only negative exchange parity states in the basis (symmetrized $(0,2)^-$ and $(4,6)^-$ and their corresponding transition matrix element). This property leads to a significant reduction of numerical cost during the calculations of collision. Although two runs are needed in the case of indistinguishable molecules, the cost of each such run is much cheaper than in the case of distinguishable molecules, because it scales polynomially with respect to the size of rotational basis set. For example, it was found that the cost of AT-MQCT calculations (where AT stands for adiabatic trajectory approximation)[31,39] scales about quadratically with respect to the basis set size. Therefore, reducing the basis set size by two, and running two independent calculations is overall cheaper, by about a factor of two.

## II-C. State-to-state transition cross sections

We outline the case of *distinguishable* molecules first. In MQCT, as in many other time-dependent methods, the collision starts with the wavefunction set up to represent a chosen initial state of the system $|j_1 j_2 j m\rangle$ and the evolution of probability amplitudes $a(t)$ for various final states $|j'_1 j'_2 j' m'\rangle$ is computed during the course of collision, for various values of the molecule-molecule orbital angular momentum $\ell$ that correlates with collision impact parameter and is varied from zero to some $\ell_{\max}$ (which is a convergence parameter). At the final moment of time, these probability amplitudes are used to compute opacity functions for all final channels $(j'_1 j'_2)$ as follows:

$$P^{(\ell)}_{j_1 j_2 j m \to j'_1 j'_2} = \sum_{j'=|j'_1-j'_2|}^{j'_1+j'_2} \sum_{m'=-j'}^{+j'} \left|a^{(\ell)}_{j'_1 j'_2 j' m'}\right|^2 \qquad (6)$$

The double sum in Eq. (6) covers all degenerate final states within the final channel, namely, all $j'$ states possible for a chosen pair of $(j'_1 j'_2)$, and all $m'$ states within these values of $j'$. Using these opacity functions, cross sections can be computed in a straightforward way as a sum over $\ell$ (i.e., over impact parameters):

$$\sigma_{j_1 j_2 j m \to j'_1 j'_2} = \frac{\pi}{2\mu U} \sum_{\ell=0}^{\ell_{max}} (2\ell + 1)\, P^{(\ell)}_{j_1 j_2 j m \to j'_1 j'_2} \qquad (7)$$

Here $U$ is the kinetic energy of collision and $\mu$ is the molecule-molecule reduced mass. However, in order to permit a more detailed and direct comparison with full-quantum methods, in the MQCT code we first compute partial cross sections labeled by the total angular momentum $J$:

$$\sigma^{(J)}_{j_1 j_2 j m \to j'_1 j'_2} = \frac{\pi}{2\mu U} \frac{2J+1}{2j+1} \sum_{\ell=|J-j|}^{J+j} P^{(\ell)}_{j_1 j_2 j m \to j'_1 j'_2} \qquad (8)$$

and then sum them over $J$:

$$\sigma_{j_1 j_2 j m \to j'_1 j'_2} = \sum_{J=0}^{J_{max}} \sigma^{(J)}_{j_1 j_2 j m \to j'_1 j'_2} \qquad (9)$$

We showed analytically,[40] and checked by many calculations that the result of Eq. (7) is equivalent to that of Eqs. (8-9). For the elastic scattering channel, a special procedure is adopted that includes the scattering phase $\delta(\ell)$:[40]

$$\sigma_{j_1 j_2 jm \to j'_1 j'_2} = \frac{\pi}{2\mu U} \sum_{\ell=0}^{\ell_{max}} (2\ell + 1) \left[ 1 - \sqrt{P^{(\ell)}_{j_1 j_2 jm \to j'_1 j'_2}} \, e^{i\delta(\ell)} \right]^2 \quad (10)$$

which will not be emphasized here. Finally, the cross section is averaged over all degenerate states $j$ and $m$ possible within the initial channel $(j_1 j_2)$:

$$\sigma_{j_1 j_2 \to j'_1 j'_2} = \frac{1}{(2j_1 + 1)(2j_2 + 1)} \sum_{j=|j_1-j_2|}^{j_1+j_2} \sum_{m=-j}^{+j} \sigma_{j_1 j_2 jm \to j'_1 j'_2} \quad (11)$$

Since MQCT trajectories are identical for positive and negative values of $m$, the code propagates trajectories for non-negative values of $m$ only, and then computes the second sum in Eq. (11) through a reduced range as

$$\sum_{m=-j}^{+j} \sigma_{j_1 j_2 jm \to j'_1 j'_2} = \sum_{m=0}^{+j} \frac{2}{1 + \delta_{m0}} \sigma_{j_1 j_2 jm \to j'_1 j'_2}$$

The denominator of Eq. (11) is computed in MQCT code as $(j_{max} + j_{min} + 1)(j_{max} - j_{min} + 1)$, where $j_{max} = j_1 + j_2$ and $j_{min} = |j_1 - j_2|$. Finally, the Billing scaling factor $(U/E)$ is applied to rescale cross section,[41] as described in detail elsewhere.[42,43] Equations similar to (6)-(11) and a discrete sampling of angular momenta for classical calculations have been derived and discussed previously for atom-diatom collisions.[44]

The expressions given above are used in the regular MQCT calculations, when the trajectories are sampled and propagated for all values of $j$, $m$ and $\ell$. Another option in the MQCT code is to sample $N$ trajectories randomly, in which case an expression for Monte-Carlo cross section can be obtained from Eqs. (6, 7, 11):

$$\sigma_{j_1 j_2 \to j'_1 j'_2} = \frac{\pi}{2\mu U} \frac{\ell_{max} + 1}{N} \sum_{i=1}^{N} w^{(i)} (2\ell + 1) P^{(\ell)}_{j_1 j_2 jm \to j'_1 j'_2} \quad (12)$$

Here the sum is over randomly sampled trajectories (few hundred is typically enough) and $w^{(i)}$ is the weight of each trajectory, used to count the number of random hits that produce the same trajectory. Again, only the trajectories with non-negative values of $m$ are sampled, but the contribution of negative values is taken into account by making the "bins" for positive $m$ twice larger than that for $m = 0$, which gives larger values of their sampling weights $w^{(i)}$.

## II-D. Cross sections for indistinguishable collision partners

Considering the fact that for identical collision partners there are two exchange parities that determine whether the transition is allowed or not, cross section should be calculated using two terms responsible for two exchange parities:[20,22]

$$\sigma_{j_1 j_2 jm \to j'_1 j'_2} = w^+ \sigma^+_{j_1 j_2 jm \to j'_1 j'_2} + w^- \sigma^-_{j_1 j_2 jm \to j'_1 j'_2} \tag{13}$$

Here the first term corresponds to transitions within positive exchange parity states while the second term represents transitions between states with negative exchange parity. The factors $w^+$ and $w^-$ are statistical weights calculated using nuclear spin $I$ of the molecule: $w^+ = (I+1)/(2I+1)$, $w^- = I/(2I+1)$.[22] Note that for *para*-H$_2$, CO and *para*-H$_2$O considered in this work the total nuclear spin $I = 0$, and therefore we have $w^+ = 1$ and $w^- = 0$.

It should also be noted that in the full-quantum calculations for identical collision partners the value of orbital angular momentum $\ell$ affects matrix elements,[20,22] introducing an additional factor $(-1)^\ell$ into the second term of the quantum analogue of Eq. (5). Due to this, the coupling between states of given exchange parity is introduced either only by *even* or only by *odd* values of $\ell$ (depending on the value of inversion parity $p$ of the state). This effect does not occur in the MQCT matrix elements, because here the scattering process is treated classically, but it can easily be taken into account during the calculation of MQCT cross sections, as follows. After the first run, done with positive exchange parity states, with matrix elements $\langle j'_1 j'_2 j'm | V(R, \Lambda_1, \Lambda_2) | j_1 j_2 jm \rangle^+$ and for all values of $\ell$, two cross sections are computed: $\sigma^{+(\text{evn})}_{j_1 j_2 jm \to j'_1 j'_2}$ and $\sigma^{-(\text{odd})}_{j_1 j_2 jm \to j'_1 j'_2}$ using only trajectories with *even* and only trajectories with *odd* values of $\ell$, respectively. After the second run, done with negative exchange parity states, with matrix elements $\langle j'_1 j'_2 j'm | V(R, \Lambda_1, \Lambda_2) | j_1 j_2 jm \rangle^-$ and for all values of $\ell$, two more cross sections are computed: $\sigma^{-(\text{evn})}_{j_1 j_2 jm \to j'_1 j'_2}$ and $\sigma^{+(\text{odd})}_{j_1 j_2 jm \to j'_1 j'_2}$, again, using only trajectories with *even* and only trajectories with *odd* values of $\ell$, respectively. These four cross sections can be combined into:

$$\sigma^+_{j_1 j_2 jm \to j'_1 j'_2} = \sigma^{+(\text{evn})}_{j_1 j_2 jm \to j'_1 j'_2} + \sigma^{+(\text{odd})}_{j_1 j_2 jm \to j'_1 j'_2} \tag{14a}$$

and

$$\sigma^-_{j_1 j_2 jm \to j'_1 j'_2} = \sigma^{-(\text{odd})}_{j_1 j_2 jm \to j'_1 j'_2} + \sigma^{-(\text{evn})}_{j_1 j_2 jm \to j'_1 j'_2} \tag{14b}$$

that can be substituted into Eq. (13). In the code, these contributions are simply accumulated over the two runs for *indistinguishable* collision partners using appropriate statistical weights, as follows:

$$\sigma_{j_1 j_2 jm \to j_1' j_2'} = w^+ \sigma^{+(evn)}_{j_1 j_2 jm \to j_1' j_2'} + w^- \sigma^{-(odd)}_{j_1 j_2 jm \to j_1' j_2'}$$
$$+ w^- \sigma^{-(evn)}_{j_1 j_2 jm \to j_1' j_2'} + w^+ \sigma^{+(odd)}_{j_1 j_2 jm \to j_1' j_2'} \quad (15)$$

It is easy to check that this method gives the desired result. It is also important to note that if one of the weights $w^+$ or $w^-$ is equal to zero, then we need only one half of trajectories, which offers another factor of two advantage for the treatment of identical collision partners as indistinguishable. For example, for all molecular systems considered here $w^- = 0$, which means that there is no need to propagate trajectories with odd values of $\ell$ for the initial states of positive exchange parity, and no need to propagate trajectories with even values of $\ell$ for the initial states of negative exchange parity (because their contributions would anyway be multiplied by $w^- = 0$).

In the Monte Carlo calculations for *indistinguishable* collision partners the same idea is implemented by random sampling of exchange parity ($\pm$) simultaneously with random sampling of the quantum numbers $j$, $m$ and $\ell$, and introducing statistical weight into Eq. (12) for cross section:

$$\sigma_{j_1 j_2 \to j_1' j_2'} = \frac{\pi}{2\mu U} \frac{\ell_{max} + 1}{N} \sum_{i=1}^{N} w^\pm w^{(i)} (2\ell + 1) P^{(\ell)}_{j_1 j_2 jm \to j_1' j_2'} \quad (16)$$

The value of $w^\pm$ here is either $w^+$ or $w^-$, depending on the combination of initial state's exchange parity ($\pm$) and the value of $\ell$ for a given randomly generated trajectory, as summarized in Table 1.

**Table 1**: The dependence of statistical weight $w^\pm$ on initial conditions in Monte-Carlo MQCT calculations.

| Exchange parity of initial state ($\pm$) | Orbital angular momentum, $\ell$ | Statistical weight for Eq. (16) |
|---|---|---|
| + | Even | $w^+$ |
| + | Odd | $w^-$ |
| − | Even | $w^-$ |
| − | Odd | $w^+$ |

Finally, in the case of indistinguishable collision partners, the cross section computed using regular sampling of trajectories, Eq. (15), or the Monte-Carlo cross section computed using random sampling, Eq. (16), either must be multiplied by a factor of $(1 + \delta_{j_1 j_2})(1 + \delta_{j'_1 j'_2})$ that originates in the quantum treatment of scattering.[24] This factor, properly included in the latest update of the MQCT code, leaves unchanged the values of cross sections for non-pair ↔ non-pair transitions, while the values of cross sections for pair ↔ non-pair transitions are doubled, and the values of cross sections for pair ↔ pair transitions are multiplied by four.

**II-E. Differences and similarities between the two methods**

Table 2 summarizes all differences between theories for distinguishable and indistinguishable collision partners and explains how one can use a conceptually simpler distinguishable approach in the case of identical collision partners to approximately match the results of rigorous but more involved method (where the collision partners are treated as indistinguishable). Here $M_{nn'}$ is a shortened notation for matrix element of $n \rightarrow n'$ transition, where $n$ represents a set of quantum numbers for the initial state $|j_1 j_2 jm\rangle$, $n'$ corresponds to the final state $\langle j'_1 j'_2 j'm|$, while $\tilde{n}$ is a set of swapped quantum numbers $|j_2 j_1 jm\rangle$.

In the simplest case of pair → pair transition, given in the first row of Table 2, only one matrix element is computed in the case of distinguishable molecules, and exactly the same matrix element is obtained from Eq. (5) in the case of indistinguishable molecules. Recall that in this case only one exchange parity is possible (only "+" or only "−", depending on inversion parity $p$, that in turn depends on the value of $j$). The factor $(1 + \delta_{j_1 j_2})(1 + \delta_{j'_1 j'_2}) = 4$ is applied, but since only one exchange parity is possible, the trajectories are propagated for one half values of $\ell$ (only even or only odd, depending on exchange parity). Therefore, the last column of Table 2 indicates that, in order to match the results of indistinguishable approach, the cross section obtained from the distinguishable calculations must be multiplied by 2.

Next row in Table 2 deals with pair → non-pair transition, in which case, again, two matrix elements in the numerator of Eq. (5) are equivalent, therefore, only one parity is present in the calculations for indistinguishable partners. If two identical molecules are treated as distinguishable, the two corresponding matrix elements are equal (as indicated in fifth column of

Table 2) and give the same cross sections. The total cross section can be computed as a sum of these (see the last column of Table 2). For example, if two identical molecules are treated as distinguishable, then cross sections for transitions $(2,2) \rightarrow (4,6)$ and $(2,2) \rightarrow (6,4)$ will be the same and should be added together to obtain cross section comparable to that of the indistinguishable treatment (or, equivalently, one of them should be multiplied by 2). This analysis implies that transition probability, and the resultant cross section, are determined by the square of transition matrix element.

**Table 2**: Comparison of matrix elements and cross sections for identical molecules treated as distinguishable and indistinguishable collision partners.

| Transitions | Indistinguishable | | | | Distinguishable | |
|---|---|---|---|---|---|---|
| Pair to Pair | $\dfrac{M_{nn'} \times 2}{2}$, one exchange parity | $\delta_{\text{factor}} = 4$ | $\dfrac{1}{2}$ of $\ell$ | | One $M_{nn'}$ | $\sigma_{nn'} \times 2$ |
| Pair to Non-Pair | $\dfrac{M_{nn'} \times 2}{\sqrt{2}}$, one exchange parity | $\delta_{\text{factor}} = 2$ | $\dfrac{1}{2}$ of $\ell$ | | $M_{nn'} = M_{n\tilde{n}'}$ | $\sigma_{nn'} + \sigma_{n\tilde{n}'}$ (or $\sigma_{nn'} \times 2$) |
| Non-Pair to Pair | $\dfrac{M_{nn'} \times 2}{\sqrt{2}}$, one exchange parity | $\delta_{\text{factor}} = 2$ | $\dfrac{1}{2}$ of $\ell$ | | $M_{nn'} = M_{\tilde{n}n'}$ | $\sigma_{nn'} \times 2$ (or $\sigma_{nn'} + \sigma_{\tilde{n}n'}$) |
| Non-Pair to Non-Pair | $M^{\pm} = M_{nn'} \pm p\, M_{\tilde{n}n'}$ two exchange parities | $\delta_{\text{factor}} = 1$ | all $\ell$ | | $M_{nn'} \neq M_{n\tilde{n}'}$ | $\sigma_{nn'} + \sigma_{n\tilde{n}'}$ |

Similar, for non-pair → pair transition, described in the third row of Table 2, only one exchange parity exists (that of the final state), two matrix elements and two cross sections are equivalent (e.g., those for transitions $(4,6) \rightarrow (2,2)$ and $(6,4) \rightarrow (2,2)$ in the example discussed above) and an approximation to the total cross section can be obtained as one of them multiplied by 2 (or as a sum of two cross sections), as indicated in the last column of Table 2. Note that since only one exchange parity is allowed, the number of trajectories is reduced by a half, keeping only even or only odd values of $\ell$ (depending on $j'$).

Only in the case of non-pair → non-pair transition described in the last row of Table 2, two exchange parities exist for each state in the indistinguishable treatment and two matrix elements are different in the distinguishable treatment leading to two different cross sections (*e.g.*, for transitions (0,2) → (4,6) and (0,2) → (6,4) in our example). These must be added together to obtain approximation to the total cross section of the indistinguishable case (see last column of Table 2).

Overall, from the last column of Table 2 it follows that, all cross sections obtained from the treatment of collision partners as distinguishable are roughly twice smaller (compared to those obtained from the rigorous treatment of identical collision partners) and therefore they need to be either multiplied by 2 or summed over physically indistinguishable transitions. A similar strategy was used in the recent paper on CO + CO collisions.[28] This simple prescription neglects quantum interference effects but is expected to be reasonably accurate in the limit of high collision energies. This hypothesis is tested by calculations in the next section.

## III. RESULTS OF CALCULATIONS

In order to check that our equations and the code are correct, we, first of all, carried out calculations for a simple well-studied system $H_2 + H_2$. For this case, state-to-state transition cross sections are available from calculations carried out by three different groups, computed using different codes, and different variations of a full-quantum scattering approach.[2,5,6] Importantly, those three sets of results are in good agreement with each other and therefore represent a reliable benchmark. In all cases the two $H_2$ molecules were treated as indistinguishable. In Fig. 1 we compare cross sections for transitions (0,0) → (2,0), (2,2), (4,0), (4,2) and (4,4) in $H_2 + H_2$ system, as a function of collision energy, obtained from our MQCT calculations to that from Lee *et al.*[2] Solid green lines were obtained in a direct way, using the ground state (0,0) as the initial state in MQCT calculations, while dashed red lines were obtained "in reverse", running five calculations with the initial states (2,0), (2,2), (4,0), (4,2) and (4,4), taking cross sections for their quenching onto the ground state (0,0), and then using the principle of microscopic reversibility to obtain the excitation cross sections.[30,42,43] Therefore, Fig. 1 covers ten state-to-state transition processes among which four are pair → pair, three are pair → non-pair and three are non-pair → pair. We see a reasonably good agreement in all cases. The differences between MQCT and full-quantum calculations are small, much smaller than would be introduced by a missing factor of 2, indicating that all symmetry related factors were taken into account correctly.

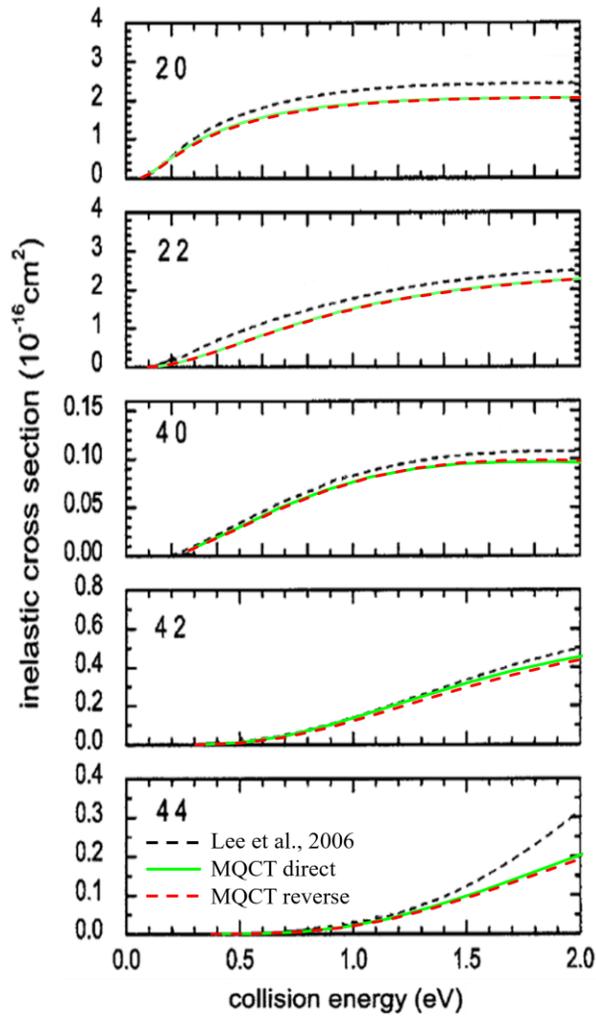

**Figure 1:** Inelastic cross sections for transitions from ground state (0,0) to five lowest excited states in $H_2 + H_2$ system. Black dashed lines correspond to full quantum data from Lee *et al.*[2] Green solid line is CC-MQCT results obtained from direct calculations, while red dashed lines are results obtained in "reverse" using microscopic reversibility.

In Fig. 2 some of these transitions are shown in the low energy range, where the results of Gatti *et al*[5] and Lin and Guo[6] are also available. We note that all four calculations were done using the same potential energy surface, but three full quantum calculations were done using slightly different methods. Namely, Lee *et al* used a time independent coupled-channel (CC) method, Gatti *et al* used a time dependent wave-packet method, while Lin and Guo used a time dependent method with Coriolis coupling neglected (CS wave-packet). The agreement between MQCT results and the results of three full-quantum methods for $H_2 + H_2$ is good without any empirical adjustments. Some differences at higher energies are likely to be due to different basis sizes. This level of accuracy is typical to the MQCT method. A similar level of agreement was found in our recent calculations for $CO + CO$,[45] treated there as distinguishable collision partners in both MQCT and full-quantum calculations.[28]

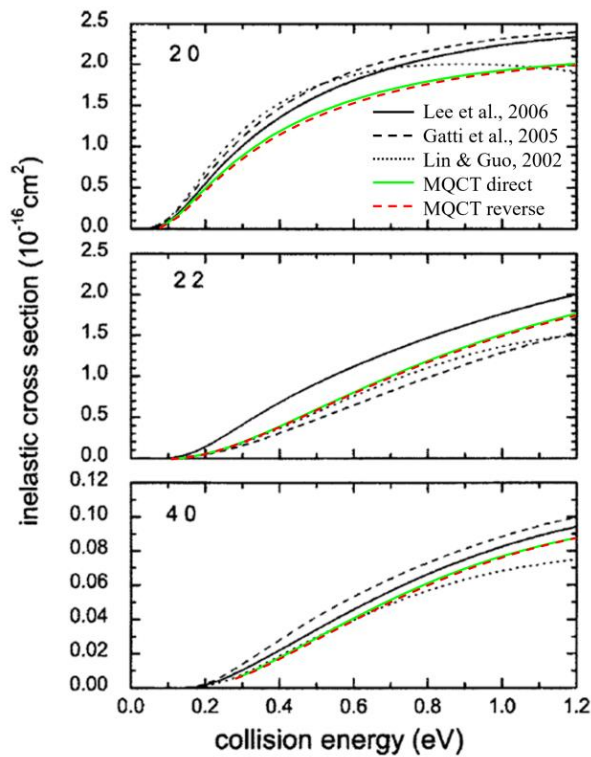

**Figure 2:** Inelastic cross sections for transitions from ground state (0,0) to three lowest excited states in $H_2 + H_2$ system. Black solid lines correspond to full quantum data from Lee *et al*,[2] black dashed lines are the results of Gatti *et al*[5] and black dotted lines are those of Lin and Guo.[6] Green solid lines are CC-MQCT results obtained from direct calculations, while red dashed lines are results obtained in "reverse" using microscopic reversibility.

In Fig. 3 we compare the results of two MQCT calculations, one in which the two $H_2$ molecules were treated as indistinguishable (same as in Figs. 1 and 2) and the other where the two $H_2$ molecules were treated as distinguishable, using a simplified approach outlined in Table 2 above. Two upper frames of Fig. 3 give examples of pair → pair transitions, two frames in the middle of Fig. 3 are examples of pair → non-pair transitions, and two frames at the bottom are non-pair → non-pair transitions. We see that the differences are very small, hard to identify on the scale of Fig. 3. It is important to note that the collision energy here is quite large (2 eV ~ 16,130 cm$^{-1}$), considering a relatively weak interaction of hydrogen molecules (~ 25 cm$^{-1}$ in the attractive well). This explains why the identical molecule exchange symmetry, being a weak quantum effect, plays very little role in these calculations.

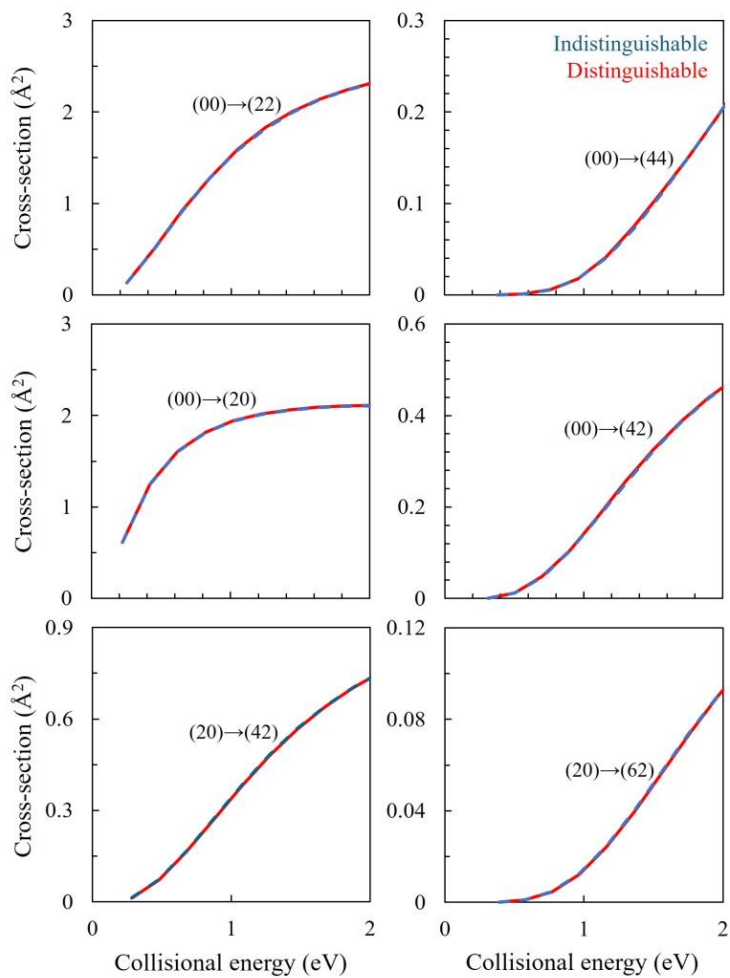

**Figure 3:** Inelastic cross sections for $H_2 + H_2$ system obtained by CC-MQCT calculations. Solid red lines correspond to the treatment of collision partners as distinguishable, while dashed blue lines are results of rigorous treatment of identical collision partners as indistinguishable.

To make small differences between the results of distinguishable and indistinguishable treatments easier to see in the figures, we computed the ratio of cross sections obtained from the two treatments (indistinguishable over distinguishable) and plotted this ratio as a function of collision energy. Using the prescription of the last column in Table 2, we included in the factor of two (or summation over indistinguishable channels) as appropriate, to ensure that in the high energy limit this ratio approaches one. All results for $H_2 + H_2$, $CO + CO$ and $H_2O + H_2O$ are presented in Figs. S1, S2 and S3 of Supplemental Information and the readers are encouraged to inspect these data. For $H_2 + H_2$ we found that the differences between distinguishable and indistinguishable treatments are negligibly small for pair ↔ pair transitions, are within few percent for non-pair ↔ non-pair transitions and are somewhat larger for pair ↔ non-pair transitions al low collision energies, but even in that case the difference is relatively small, within 7% of cross section values (see Fig. S1). Moreover, as collision energy is raised to $E \sim 16{,}130$ cm$^{-1}$ this difference monotonically decreases to ~ 3%.

For $CO + CO$ and $H_2O + H_2O$ the behavior is overall similar to that for $H_2 + H_2$, but several important differences are worth mentioning. First of all, the dipole-dipole interactions in these systems lead to deeper potential energy wells, capable of supporting multiple scattering resonances at low collision energies. The positions of these resonances are slightly different in the cases of indistinguishable and distinguishable treatments, which results in much larger (accidental) differences between the two and leads to non-monotonic behavior of the ratio. This is observed near collision energy $E \sim 100$ cm$^{-1}$ in the case of $CO + CO$ (Fig. S2) and up to $E \sim 500$ cm$^{-1}$ in the case of $H_2O + H_2O$ (Fig. S3). In this paper we will not focus on low energy resonant features, since the range of MQCT applications is at higher energies where the relative motion of collision partners can be described classically. Importantly, we found that as the collision energy is increased above resonances, the ratio of cross sections obtained by indistinguishable and distinguishable treatments approaches unity for both $CO + CO$ and $H_2O + H_2O$ systems, just like in the case of $H_2 + H_2$. A typical example is given in Fig. 4, where we present some of our results for $CO + CO$ system including 12 different pair ↔ pair transitions in the upper frame, 45 different pair ↔ non-pair transitions in the middle frame, and 27 different non-pair ↔ non-pair transitions in the lower frame. The states of CO molecule up to $j = 4$ are included.

In Fig. 4 we see that for pair ↔ pair transitions (upper frame) the differences between indistinguishable and distinguishable treatments of $CO + CO$ are very small for all energies above the scattering resonances. For pair ↔ non-pair and non-pair ↔ non-pair transitions the differences between indistinguishable and distinguishable treatments of $CO + CO$ are visible, but remain relatively small, within 5% for the majority of considered transitions and they tend to decrease as collision energy is raised. There are, however, several transitions where the differences up to 14% were observed (indicated in Fig. 4 by red labels). The origin of these differences was investigated in detail and is presented in the next section.

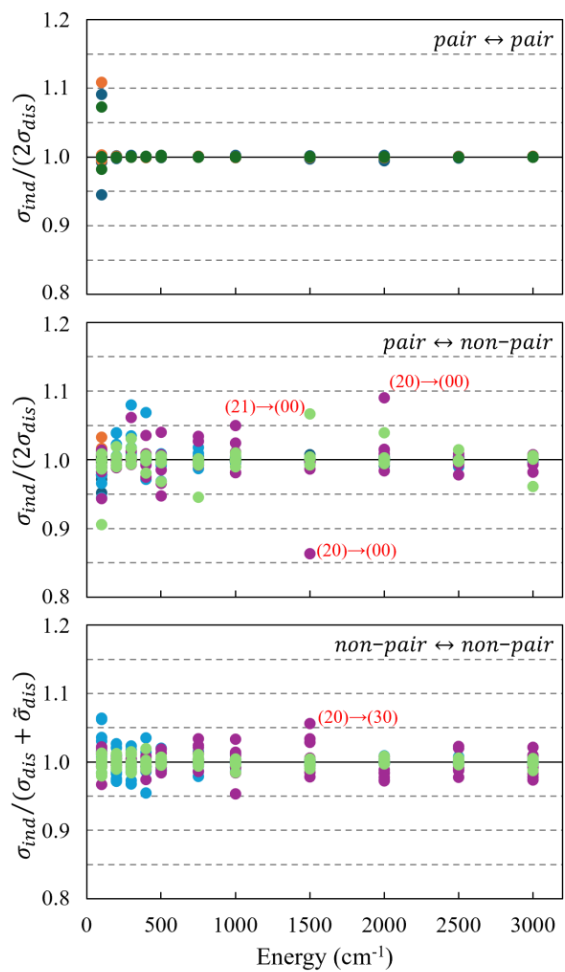

**Figure 4:** Ratios of inelastic cross sections for CO + CO obtained by treating collision partners either as indistinguishable or distinguishable in CC-MQCT calculations at different collision energies. Top frame displays data for pair ↔ pair transition, middle frame for pair ↔ non-pair and lower frame for non-pair ↔ non-pair transitions. Transitions are grouped by their initial states $(j_1 j_2)$ and each initial state is assigned a certain color of symbols: blue for (00), orange for (22), dark green for (44), light blue for (20), violet for (40), and light green for (42). A complete list of included transitions is given in Table S3 of Supplemental Information. Several transitions with large deviations of the ratio from unity are labelled. Some of them are further explored in Fig. 5.

## IV. DISCUSSION

The results presented in Figs. 3-4 above, and in Figs. S1-S3 of the Supplemental Information indicate that, overall, the treatments of identical collision partners as distinguishable or indistinguishable lead to very similar results (besides the factor of 2 that can easily be taken into account *a posteriori*). The differences between the two methods can be larger in the low-energy regime dominated by scattering resonances, but at higher energies the differences are typically within 5% and tend to decrease with increasing collision energy. Some transitions show larger deviations, and we used these cases to trace the origin of differences. We investigated several of these transitions and present three examples here. The first is (20) → (00) transition in CO + CO that belongs to non-pair → pair type and exhibits 14% difference between indistinguishable and distinguishable treatments when the collision energy is 1500 cm$^{-1}$. For this process in Fig. 5 we plotted partial cross sections (as a function of orbital angular momentum quantum number $\ell$) obtained using the two treatments of CO + CO partners. Three frames of the figure correspond to three $(jm)$ components of the initial state indicated as superscript: $(j_1 j_2)^{jm} = (20)^{20}, (20)^{21}, (20)^{22}$. The factor of 2, or summation over physically indistinguishable transitions, has already been applied prior to plotting Fig. 5, to correct the results of distinguishable treatment. From Fig. 5 we clearly see that for large values of impact parameter, $\ell > 150$, the two treatments give nearly identical results. For the intermediate range of impact parameter, $100 > \ell > 150$, the dependencies of partial cross section look very similar for the indistinguishable and distinguishable treatments and show all the same features of the transition but appear slightly *shifted* one with respect to another. For even smaller impact parameters, $\ell < 100$, this shift increases and eventually leads to a different behavior of cross sections near $\ell \sim 50$. In this range, partial cross sections for the $(20)^{20}$ component of the initial state (upper frame of Fig. 5) are quite different in the two methods, producing the aforementioned difference of 14% in the total cross section.

The second example is (20) → (30) transition in CO + CO that belongs to non-pair → non-pair type and exhibits 6% difference between indistinguishable and distinguishable treatments when the collision energy is 1500 cm$^{-1}$. For this process we plotted partial cross sections in Fig. 6. Again, we clearly see that for $\ell > 150$ the two treatments give nearly identical results but appear *shifted* one with respect to another for smaller impact parameters. The major difference comes from $\ell \sim 50$ and mostly from the $(20)^{20}$ component of the initial state (upper frame of Fig. 6), which explains the difference of total cross sections in the two methods.

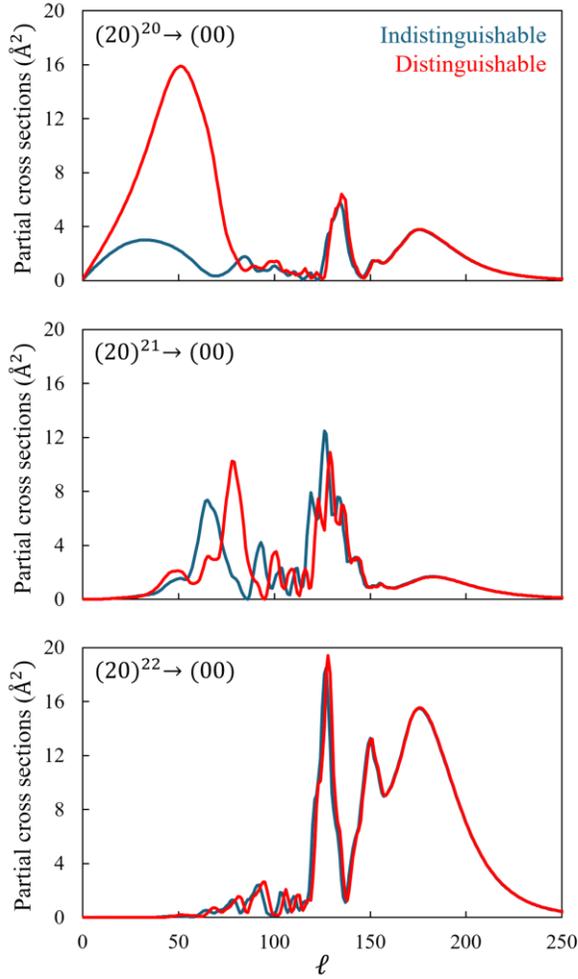

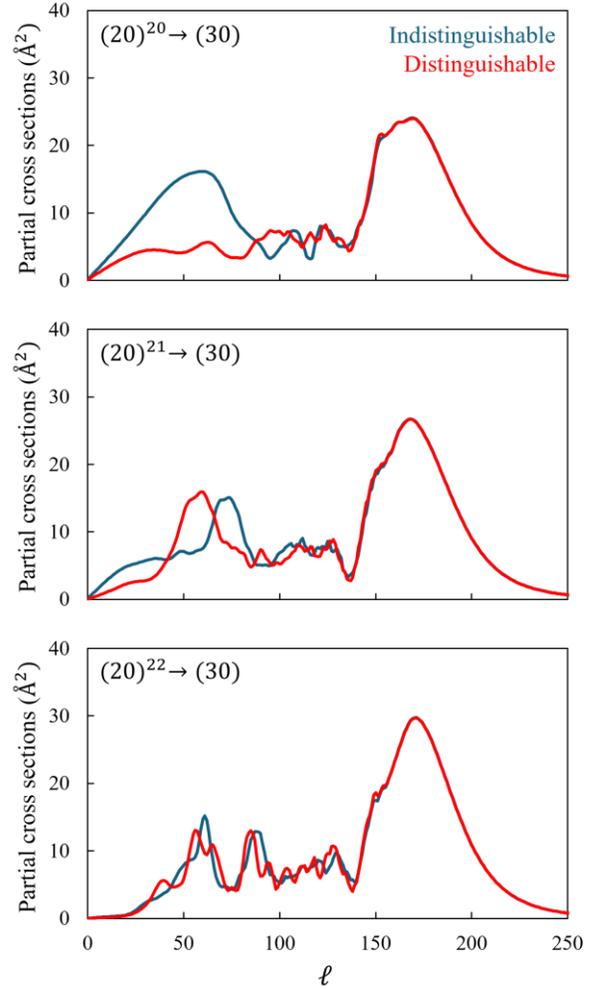

**Figure 5:** Comparison of partial cross-sections for transition $(20) \rightarrow (00)$ of non-pair $\rightarrow$ pair type, obtained using two treatments of CO + CO partners: indistinguishable (blue lines) and distinguishable (red lines) at collision energy 1500 cm$^{-1}$. Three frames correspond to three $jm$ components of the initial state, labelled as $(20)^{jm}$ in the figure. CC-MQCT method is used.

**Figure 6:** The same as Figure 5, but for transition $(20) \rightarrow (30)$ that belongs to non-pair $\rightarrow$ non-pair type.

Finally, in Fig. 7 we plotted the results for $(1_{11}0_{00}) \to (2_{11}2_{11})$ transition in $H_2O + H_2O$ system at collision energy 1000 cm$^{-1}$. For this transition of non-pair → pair type the difference of total cross sections obtained by the two treatments is more significant, about 20%. Again, we clearly see that this difference comes mostly from partial cross sections with smaller impact parameters, $\ell < 100$, and the results of two treatments appear *shifted* one with respect to another.

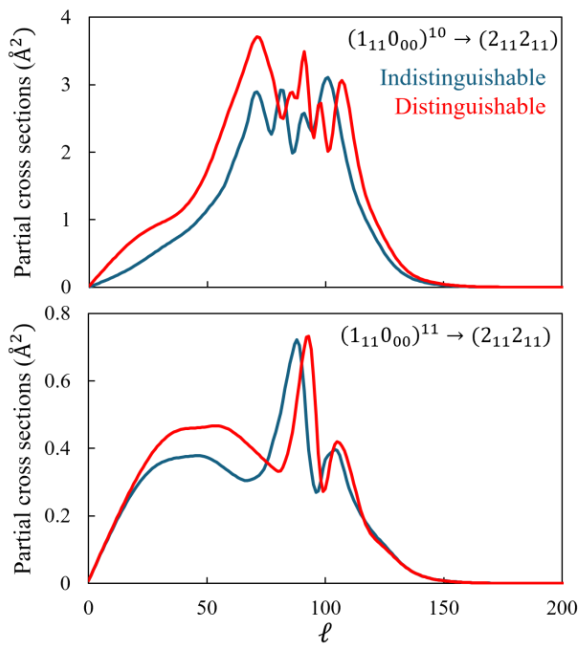

**Figure 7:** Comparison of partial cross-sections for transition $(1_{11}0_{00}) \to (2_{11}2_{11})$ of non-pair → pair type, obtained using two treatments of $H_2O + H_2O$ partners: indistinguishable (blue lines) and distinguishable (red lines) at collision energy 1000 cm$^{-1}$. Two frames correspond to two $jm$ components of the initial state, labelled as $(1_{11}0_{00})^{jm}$ in the figure. CC-MQCT method is used.

From these three examples, and several other transitions we investigated in detail in $H_2$ + $H_2$, CO + CO and $H_2O + H_2O$ systems, we came up with two conclusions. First, it becomes clear that indistinguishable and distinguishable treatments of collision partners explore the same landscape of interaction, but in slightly different ways, which manifests as a shift of partial cross section dependencies (transition probabilities) with respect to the orbital angular momentum $\ell$ (or collision impact parameter). The contributions of trajectories with large values of $\ell$ are essentially the same in the two treatments, which makes the overall results very similar. Importantly, the differences between the two treatments originate in the trajectories with small values of $\ell$ that correspond to small impact parameters, strong encounters and significant deflections of collision partners.

Still, concerning those MQCT trajectories where the results of indistinguishable and distinguishable treatments are different (at small values of $\ell$), we would like to understand what exactly makes them different. In Fig. 8 we plotted the off-diagonal (inelastic) and diagonal (elastic) matrix elements for indistinguishable and distinguishable treatments of $(20) \to (30)$ transition in CO + CO investigated in Fig. 6 above. First consider the behavior of the off-diagonal matrix elements (upper frame of Fig. 8). In the treatment of collision partners as distinguishable they describe (20) → (30) and (02) → (30) transitions, red curves in Fig. 8. In the treatment of collision partners as indistinguishable they describe $(20)^+ \to (30)^+$ and $(20)^- \to (30)^-$ transitions, blue curves in Fig. 8. One can see that, as a function of $R$, all four off-diagonal matrix elements behave quite differently, and it is very tempting to say that these differences are responsible for small differences between inelastic cross sections obtained by indistinguishable and distinguishable treatments. However, one must realize that different rotational basis sets in the indistinguishable and distinguishable treatments (of the same physical system of identical collision partners) are mathematically equivalent and although the off-diagonal matrix elements look very different in those two treatments, we found that they would produce *exactly the same inelastic transitions* if the system were carried through the *same path* during the collision process (as we will demonstrate further below by a numerical experiment illustrated by Figs. 9 and 10). The treatment of identical collision partners as indistinguishable gives us, merely, a more convenient and economic basis set for the description of inelastic transitions, but the two basis sets are, actually, equivalent.

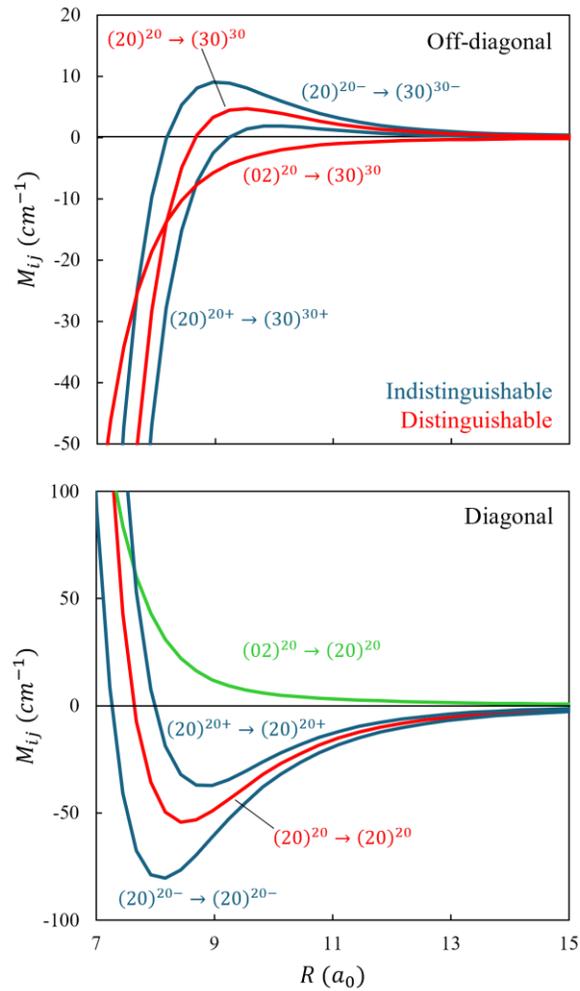

**Figure 8:** Upper frame: Off-diagonal matrix elements for $(20) \to (30)$ transition in CO + CO described as indistinguishable (blue lines) or distinguishable (red lines) collision partners. Two red curves are combined to obtain two blue curves according to Eq. (5). Lower frame: Diagonal matrix elements for the same initial state described as indistinguishable (blue lines) or distinguishable (red and green lines) collision partners. Red and green curves are combined to obtain two blue curves according to Eq. (5).

Now let's look at the behavior of the diagonal matrix elements (lower frame of Fig. 8). For the treatment of collision partners as distinguishable they describe (20) → (20) and (02) → (02) processes (red) that overlap in the figure, while for the treatment of collision partners as indistinguishable they describe $(20)^+ \to (20)^+$ and $(20)^- \to (20)^-$ transitions (blue). Once again, we see that, as a function of $R$, these diagonal matrix elements behave quite differently. This is explained by the fact that in the indistinguishable case the matrix elements $(20)^\pm \to (20)^\pm$ are computed using Eq. (5) as constructive and destructive superpositions of matrix elements for transitions (20) → (20) and (02) → (20), out of which the former is a true elastic process with large probability, while the latter is actually an inelastic process with smaller probability (green curve in Fig. 8). This explains why the matrix elements for $(20)^+ \to (20)^+$ and $(20)^- \to (20)^-$ are different one from another and are different from the original (20) → (20) matrix element, that in turn is identical to (02) → (02) matrix element.

We found that differences between the two treatments come exclusively from differences in *the diagonal matrix elements*, that govern the evolution of system along the mean-field trajectories that are nearly elastic (since inelastic transition probabilities are usually small). These trajectories are somewhat different in indistinguishable and distinguishable treatments, which manifest as the "shifts" of partial cross sections in the range of small impact parameters (seen in Figs. 5, 6 and 7 for $\ell < 150$) and leads up to 20% differences of the total cross sections (seen in Fig. 4 and Figs. S1-S3). This explanation makes sense on a qualitative level, because the largest differences between the two treatments are observed at small impact parameters, when the trajectories penetrate deeper into the interaction region and the effect is expected to be larger.

To offer a rigorous quantitative proof of this explanation, we carried out an additional computational experiment using a simplified version of MQCT method called AT-MQCT, where AT stands for adiabatic trajectory. In this method,[31,39] the classical and quantum degrees of freedom are decoupled. The classical trajectories are propagated first (typically with smaller basis set, or even using only one initial state in the basis, which gives the name "adiabatic") and the trajectory information is saved to a file. Next, the quantum equations of motion are propagated (with large basis set) using information about trajectory saved during the previous run. This approximation permits to run MQCT calculations much faster,[31,39] but this is not the goal here. Here it is important that we can save the trajectory information during MQCT calculations when the two collision partners are treated as distinguishable, and then, during the second run, employ the state-to-state transition matrix either from the indistinguishable treatment, or from the distinguishable treatment. In the course of this experiment, both indistinguishable and distinguishable calculations are caried through exactly the same collision path, but with different rotational basis sets and different state-to-state transition matrices, either indistinguishable or distinguishable. The goal of this experiment is not just to break more accurate indistinguishable calculations by forcing them to evolve along the path obtained from less accurate distinguishable calculations but is rather to observe what happens if distinguishable and indistinguishable calculations (with their different states, basis sets, and state-to-state transition matrixes) are both propagated along the same one path. Note that although inelastic transitions during the scattering process are determined by off-diagonal matrix elements, the scattering path itself is determined mainly by the diagonal matrix element of the initial quantum state, simply because inelastic transition probabilities are typically small and the MQCT trajectories are very close to those of an elastic scattering process. Therefore, this computational experiment permits to identify the origin of differences between the indistinguishable and distinguishable treatments. Is it in the inelastic transition probabilities (off-diagonal elements of the matrix) or in the scattering paths (diagonal matrix elements)?

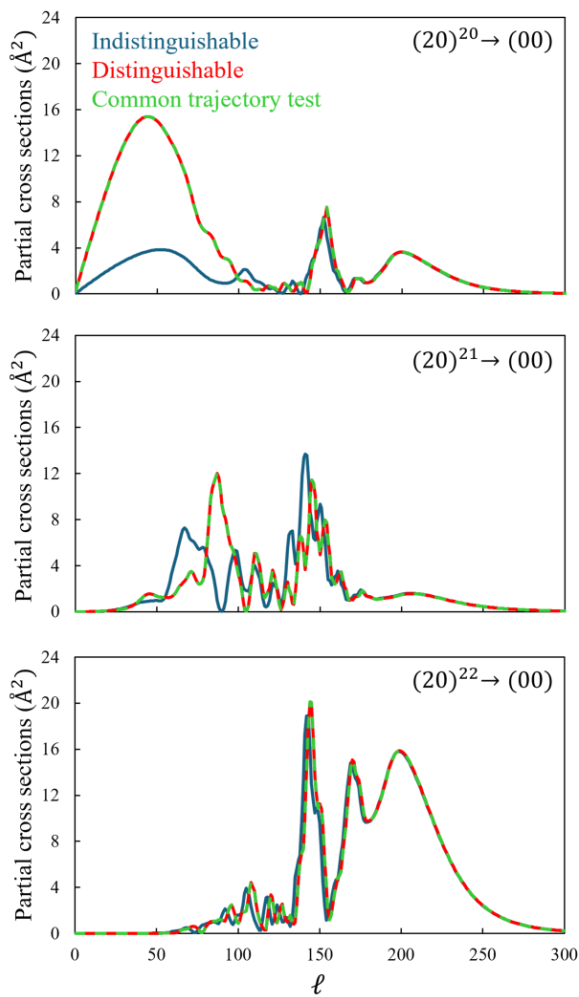
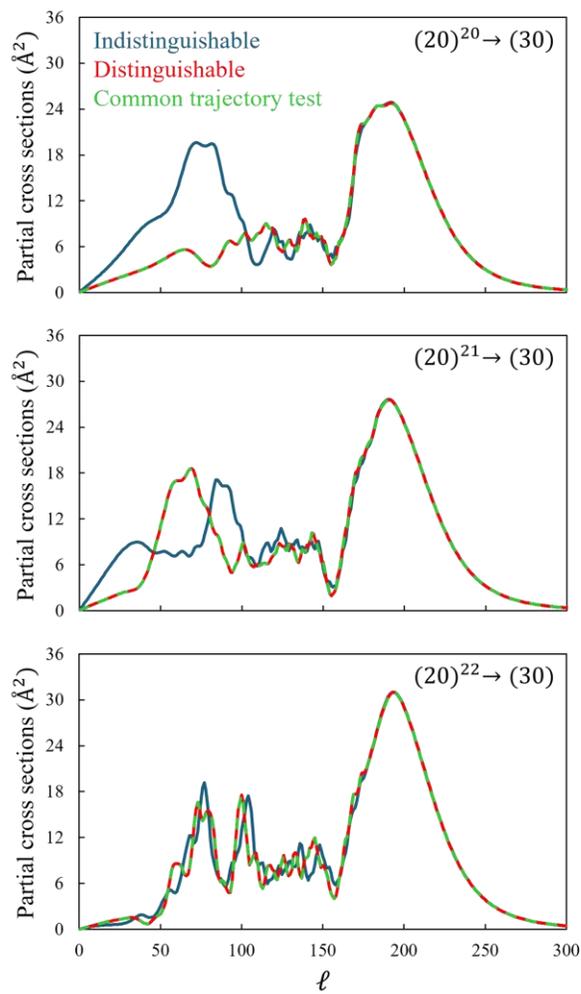

**Figure 9:** Same as Fig. 5 but now obtained using AT-MQCT method at collision energy 2000 cm$^{-1}$. Results of the common trajectory "experiment" (see text) are plotted as green dashed lines.

**Figure 10:** Same as Fig. 6 but now obtained using AT-MQCT method at collision energy 2000 cm$^{-1}$. Results of the common trajectory "experiment" (see text) are plotted as green dashed lines.

The results are presented in Figs. 9 and 10 for the same transitions that were presented in Figs. 5 and 6. As before, red and blue curves correspond to distinguishable and indistinguishable treatments, respectively, but now using AT-MQCT version of our method. The results are very similar to those presented in Figs. 5 and 6. The dashed green curves, new in Figs. 8 and 9, represent the result of the common trajectory experiment, where the indistinguishable treatment uses trajectory info from the distinguishable treatment. One can see that in this case the results of distinguishable treatment coincide almost exactly with those of indistinguishable treatment, except tiny numerical differences not visible in the figures. It is almost striking to see that the employment of the common trajectory information, basically, turns blue curves in Figs. 9 and 10 into the red ones (strictly speaking into dashed green curves, that are almost identical to the red ones).

One can also look at this computational experiment under a slightly different angle, by asking a question: If one goes from a less accurate distinguishable treatment to more accurate indistinguishable treatment, would it be sufficient to keep the same trajectories of molecule-molecule scattering and only replace the state-to-state-transition matrix with a symmetrized one (as correctly implemented in the indistinguishable treatment) hoping that it will give us different (and correct) inelastic transitions. We showed that this computational experiment does not change the results of distinguishable treatment at all! The indistinguishable treatment implemented in this way would only use a (mathematically) different basis set for solution of the same (physical) problem, which would lead to the same final result. The result different from the distinguishable treatment is obtained only when the indistinguishable treatment is applied self-consistently including its specific trajectories governed mainly by the diagonal matrix elements of the symmetrized basis set.

This numerical experiment gives a solid support to the explanation of differences found between the treatments of identical collision partners as indistinguishable or distinguishable. These differences, typically small, come from differences in the diagonal matrix elements, normally thought to affect the elastic scattering process only. However, small changes of the scattering path permit to access different parts of the interaction landscape, especially for the trajectories with small values of impact parameter (small values of orbital angular momentum $\ell$) that penetrate deeper into the repulsive area. This, in turn, leads to slightly different inelastic transitions in the cases of indistinguishable and distinguishable treatments. Symmetrization of molecular wavefunctions, and the following modification of diagonal and off-diagonal matrix elements, is responsible for this effect, but it proceeds mostly through the diagonal matrix elements, which is a somewhat unexpected phenomenon. If, artificially, the trajectories are forced to follow the same scattering path, which can be achieved using AT-MQCT version of the method, then exactly the same inelastic transitions are observed in the indistinguishable and distinguishable treatments (besides the well-known factor of 2).

## V. CONCLUSIONS

In this paper we outlined the details of mixed quantum/classical theory, MQCT, for the treatment of rotationally inelastic transitions during collisions of two identical molecules, described either as indistinguishable or distinguishable partners. A physically correct treatment of the two molecules as indistinguishable includes symmetrization of rotational wavefunctions and introduces exchange parity, which gives state-to-state transition matrix elements different from those in the straightforward treatment of molecules as distinguishable. Not only this symmetrization is physically grounded and, strictly speaking is required for the correct description of the collision process, it also carries several advantages from the practical perspective. The calculation of symmetrized matrix elements in the indistinguishable case requires smaller numerical effort and gives numerical advantage close to the factor of two. Next, the calculations of molecule + molecule collision process (here the propagation of MQCT trajectories) is also faster in the case of indistinguishable treatment due to smaller size of state-to-state transition matrixes involved, which gives another advantage close to the factor of two. These two sources of savings are general and should make the treatment of indistinguishable collision partners faster by a factor close to four for any system of two identical molecules. Moreover, for some molecules where the nuclear spin weight of one of exchange parities is zero (which is the case for $H_2 + H_2$, $CO + CO$ and $H_2O + H_2O$ systems considered here), another acceleration by a factor of two is possible, because only the trajectories with even or only with odd values of $\ell$ are required for each exchange parity. In these cases, the numerical advantage of treating the identical molecules as indistinguishable is quite substantial, close to the factor of eight. These advantages will be implemented in the next release of MQCT suite of codes.[46]

Still, many calculations in the literature, including our own recent calculations for $CO + CO$ and $H_2O + H_2O$,[45,47] were carried out treating the two molecules as distinguishable and applying an *a posteriori* correction. Sometimes this is done for historical reasons (to compare with results of an earlier work that was done in this way), and sometimes because the symmetrization of wavefunctions was not included in some codes. Therefore, we carefully compared the theory and results of indistinguishable and distinguishable treatments within MQCT framework and came up with reasonably rigorous justification of this *a posteriori* correction summarized in the last column of Table 2. It says that for pair ↔ pair transitions cross sections obtained from the distinguishable treatment need to be multiplied by 2, while for non-pair ↔ non-pair transitions they need to be summed over physically indistinguishable transitions. For pair ↔ non-pair transitions one can either apply a factor of 2, or equivalently a summation over indistinguishable transitions. This is an approximation, of course, that neglects the effect of quantum interference.

However, the calculations carried out for $H_2 + H_2$, $CO + CO$ and $H_2O + H_2O$ systems using different methods within MQCT framework, presented in the Supplemental Information, indicate that this is a very good approximation for energies above scattering resonances (those appear to be very sensitive to small details, such as symmetrization of wave functions). At higher energies the results of indistinguishable and distinguishable treatments are practically the same for all pair ↔ pair transitions, and are very similar for other types of transitions, with differences typically within 5%. Only a few individual transitions showed larger differences, within 10-20%. Also, we saw that as energy is increased, these differences monotonically decrease.

A detailed analysis of these differences showed that they originate in the diagonal elements of state-to-state transition matrix, normally thought of as something responsible for the elastic scattering processes only. However, we found that the diagonal matrix elements are slightly different in the treatments of identical collision partners as indistinguishable or distinguishable, which drives trajectories through slightly different paths in the two treatments which in turn produces slightly different cross sections for inelastic state-to-state transitions. We demonstrated that if the differences of diagonal matrix elements are neglected, by artificially dragging the system through the same path in the indistinguishable and distinguishable treatments, then all differences between the results of two treatments disappear and they become entirely equivalent.

In the future the results of this work will permit us to run more efficiently the calculations of inelastic state-to-state transition cross sections to populate databases of rate coefficients for astrochemical modeling of energy transfer in $CO + CO$ and $H_2O + H_2O$ collisions, that were treated as collisions of distinguishable partners in the past. Not only the speed up by almost an order of magnitude is possible in these cases, but the treatment of identical collision partners as indistinguishable is also expected to be more accurate, particularly at lower collision energies, compared to the employment of an approximate *a posteriori* correction applied to the results of the treatment where the two partners are treated as distinguishable (*i.e.*, factor of 2 and/or by summation over physically indistinguishable processes).

## CONFLICTS OF INTEREST

Authors declare no conflict of interests.

## DATA AVAILABILITY

The code for MQCT calculations of molecule + molecule rotationally inelastic collisions can be found at https://github.com/MarquetteQuantum/MQCT.

## AKNOWLEDGEMENTS

This research was supported by NSF grant number CHE-2102465. D. Babikov acknowledges the support of Way Klingler Research Fellowship and of the Haberman-Pfletschinger Research Fund. D. Bostan acknowledges the support of the Eisch Fellowship and the Richard W. Jobling Distinguished Fellowship. We used resources of the National Energy Research Scientific Computing Center, which is supported by the Office of Science of the U.S. Department of Energy under Contract No. DE-AC02-5CH11231. This research also used HPC resources at Marquette, funded in part by the National Science foundation award CNS-1828649.

*Supplemental Information for:*

# "On Mixed Quantum/Classical Theory for Rotationally Inelastic Scattering of Identical Collision Partners"
D. Bostan, B. Mandal and D. Babikov

In this work we did not pursue the goal of producing cross sections converged with respect to the size of rotational basis set. Therefore, small basis sets were used, sufficient to introduce all kinds of transitions as discussed in the paper: between pair ↔ pair, pair ↔ non-pair, non-pair ↔ non-pair states. These are summarized in Table S1. Other input parameters are given in Table S2.

**Table S1**: Rotational basis sets employed and the number of non-zero matrix elements for various collision partners considered in this work.

| Collision partners | Maximum $j_1, j_2$ | Distinguishable | | | Indistinguishable | | |
|---|---|---|---|---|---|---|---|
| | | Channels | $j_1 j_2 j m$ states | Matrix elements | Channels | $j_1 j_2 j m^{\pm}$ states* | Matrix elements* |
| $H_2 + H_2$ | 6 (even only) | 16 | 784 | 20404 | 10 | 406, 378 | 5505, 4911 |
| $CO + CO$ | 4 | 25 | 625 | 19395 | 15 | 325, 300 | 5249, 4626 |
| $p$-$H_2O$ + $p$-$H_2O$ | 2 | 25 | 361 | 9862 | 15 | 190, 171 | 2724, 2316 |

*First and second numbers correspond to positive and negative exchange parities, respectively.

**Table S2**: Other input parameters of MQCT calculations.

| Collision partners | $R_{min} - R_{max}$ (Bohr) | Reduced mass (amu) | Max. impact parameter (Bohr) | Number of grid points | | | |
|---|---|---|---|---|---|---|---|
| | | | | $R$ | $\alpha_1, \alpha_2$ | $\beta_1, \beta_2$ | $\gamma_1, \gamma_2$ |
| $H_2 + H_2$ | 2.0−50.0 | 1.0 | 20.0 | 100 | 0, 50 | 25, 25 | 0, 0 |
| $CO + CO$ | 5.0 −50.0 | 13.99746 | 20.0 | 76 | 0, 40 | 20, 20 | 0, 0 |
| $H_2O + H_2O$ | 4.0 −100.0 | 9.00764 | 30.0 | 100 | 30,30 | 15,15 | 30,30 |

**Table S3:** The list of transitions presented in this work for each of three systems studied. Transitions are grouped by their initial states, and each initial state is assigned a certain color. These colors are the same as colors of symbols in Fig. 4 of the main text and in Figs. S1-S3 below.

| | pair ↔ pair (6 transitions) | pair ↔ non-pair (18 transitions) | | non-pair ↔ non-pair (6 transitions) | |
|---|---|---|---|---|---|
| **H$_2$ + H$_2$** | (00) → (22) <br> (00) → (44) | (00) → (20) <br> (00) → (40) <br> (00) → (42) | (20) → (00) <br> (20) → (22) <br> (20) → (44) | (20) → (40) <br> (20) → (42) | |
| | (22) → (00) <br> (22) → (44) | (22) → (20) <br> (22) → (40) <br> (22) → (42) | (40) → (00) <br> (40) → (22) <br> (40) → (44) | (40) → (20) <br> (40) → (42) | |
| | (44) → (00) <br> (44) → (22) | (44) → (20) <br> (44) → (40) <br> (44) → (42) | (42) → (00) <br> (42) → (22) <br> (42) → (44) | (42) → (20) <br> (42) → (40) | |

| | pair ↔ pair (12 transitions) | pair ↔ non-pair (45 transitions) | | non-pair ↔ non-pair (27 transitions) | |
|---|---|---|---|---|---|
| **CO + CO** | (00) → (11) <br> (00) → (22) <br> (00) → (33) <br> (00) → (44) | (00) → (10) <br> (00) → (20) <br> (00) → (21) <br> (00) → (30) <br> (00) → (31) | (00) → (32) <br> (00) → (40) <br> (00) → (41) <br> (00) → (42) <br> (00) → (43) | (10) → (00) <br> (10) → (11) <br> (10) → (22) <br> (10) → (33) <br> (10) → (44) | (10) → (20) <br> (10) → (21) <br> (10) → (30) <br> (10) → (31) <br> (10) → (32) | (10) → (40) <br> (10) → (41) <br> (10) → (42) <br> (10) → (43) |
| | (11) → (00) <br> (11) → (22) <br> (11) → (33) <br> (11) → (44) | (11) → (10) <br> (11) → (20) <br> (11) → (21) <br> (11) → (30) <br> (11) → (31) | (11) → (32) <br> (11) → (40) <br> (11) → (41) <br> (11) → (42) <br> (11) → (43) | (20) → (00) <br> (20) → (00) <br> (20) → (22) <br> (20) → (33) <br> (20) → (44) | (20) → (10) <br> (20) → (21) <br> (20) → (30) <br> (20) → (31) <br> (20) → (32) | (20) → (40) <br> (20) → (41) <br> (20) → (42) <br> (20) → (43) |
| | (22) → (00) <br> (22) → (11) <br> (22) → (33) <br> (22) → (44) | (22) → (10) <br> (22) → (20) <br> (22) → (21) <br> (22) → (30) <br> (22) → (31) | (22) → (32) <br> (22) → (40) <br> (22) → (41) <br> (22) → (42) <br> (22) → (43) | (21) → (00) <br> (21) → (11) <br> (21) → (22) <br> (21) → (33) <br> (21) → (44) | (21) → (10) <br> (21) → (20) <br> (21) → (30) <br> (21) → (31) <br> (21) → (32) | (21) → (40) <br> (21) → (41) <br> (21) → (42) <br> (21) → (43) |

| | pair ↔ pair (9 transitions) | pair ↔ non-pair (30 transitions) | | non-pair ↔ non-pair (15 transitions) | |
|---|---|---|---|---|---|
| **H$_2$O + H$_2$O** | $(0_{00}0_{00}) \to (1_{11}1_{11})$ <br> $(0_{00}0_{00}) \to (2_{02}2_{02})$ <br> $(0_{00}0_{00}) \to (2_{11}2_{11})$ | $(0_{00}0_{00}) \to (1_{11}0_{00})$ <br> $(0_{00}0_{00}) \to (2_{02}0_{00})$ <br> $(0_{00}0_{00}) \to (2_{02}1_{11})$ <br> $(0_{00}0_{00}) \to (2_{11}0_{00})$ <br> $(0_{00}0_{00}) \to (2_{11}1_{11})$ <br> $(0_{00}0_{00}) \to (2_{11}2_{02})$ | $(1_{11}0_{00}) \to (0_{00}0_{00})$ <br> $(1_{11}0_{00}) \to (1_{11}1_{11})$ <br> $(1_{11}0_{00}) \to (2_{02}2_{02})$ <br> $(1_{11}0_{00}) \to (2_{11}2_{11})$ | $(1_{11}0_{00}) \to (2_{02}0_{00})$ <br> $(1_{11}0_{00}) \to (2_{02}1_{11})$ <br> $(1_{11}0_{00}) \to (2_{11}0_{00})$ <br> $(1_{11}0_{00}) \to (2_{11}1_{11})$ <br> $(1_{11}0_{00}) \to (2_{11}2_{02})$ | |
| | $(1_{11}1_{11}) \to (0_{00}0_{00})$ <br> $(1_{11}1_{11}) \to (2_{02}2_{02})$ <br> $(1_{11}1_{11}) \to (2_{11}2_{11})$ | $(1_{11}1_{11}) \to (1_{11}0_{00})$ <br> $(1_{11}1_{11}) \to (2_{02}0_{00})$ <br> $(1_{11}1_{11}) \to (2_{02}1_{11})$ <br> $(1_{11}1_{11}) \to (2_{11}0_{00})$ <br> $(1_{11}1_{11}) \to (2_{11}1_{11})$ <br> $(1_{11}1_{11}) \to (2_{11}2_{02})$ | $(2_{02}0_{00}) \to (0_{00}0_{00})$ <br> $(2_{02}0_{00}) \to (1_{11}1_{11})$ <br> $(2_{02}0_{00}) \to (2_{02}2_{02})$ <br> $(2_{02}0_{00}) \to (2_{11}2_{11})$ | $(2_{02}0_{00}) \to (1_{11}0_{00})$ <br> $(2_{02}0_{00}) \to (2_{02}1_{11})$ <br> $(2_{02}0_{00}) \to (2_{11}0_{00})$ <br> $(2_{02}0_{00}) \to (2_{11}1_{11})$ <br> $(2_{02}0_{00}) \to (2_{11}2_{02})$ | |
| | $(2_{02}2_{02}) \to (0_{00}0_{00})$ <br> $(2_{02}2_{02}) \to (1_{11}1_{11})$ <br> $(2_{02}2_{02}) \to (2_{11}2_{11})$ | $(2_{02}2_{02}) \to (1_{11}0_{00})$ <br> $(2_{02}2_{02}) \to (2_{02}0_{00})$ <br> $(2_{02}2_{02}) \to (2_{02}1_{11})$ <br> $(2_{02}2_{02}) \to (2_{11}0_{00})$ <br> $(2_{02}2_{02}) \to (2_{11}1_{11})$ <br> $(2_{02}2_{02}) \to (2_{11}2_{02})$ | $(2_{02}1_{11}) \to (0_{00}0_{00})$ <br> $(2_{02}1_{11}) \to (1_{11}1_{11})$ <br> $(2_{02}1_{11}) \to (2_{02}2_{02})$ <br> $(2_{02}1_{11}) \to (2_{11}2_{11})$ | $(2_{02}1_{11}) \to (1_{11}0_{00})$ <br> $(2_{02}1_{11}) \to (2_{02}0_{00})$ <br> $(2_{02}1_{11}) \to (2_{11}0_{00})$ <br> $(2_{02}1_{11}) \to (2_{11}1_{11})$ <br> $(2_{02}1_{11}) \to (2_{11}2_{02})$ | |

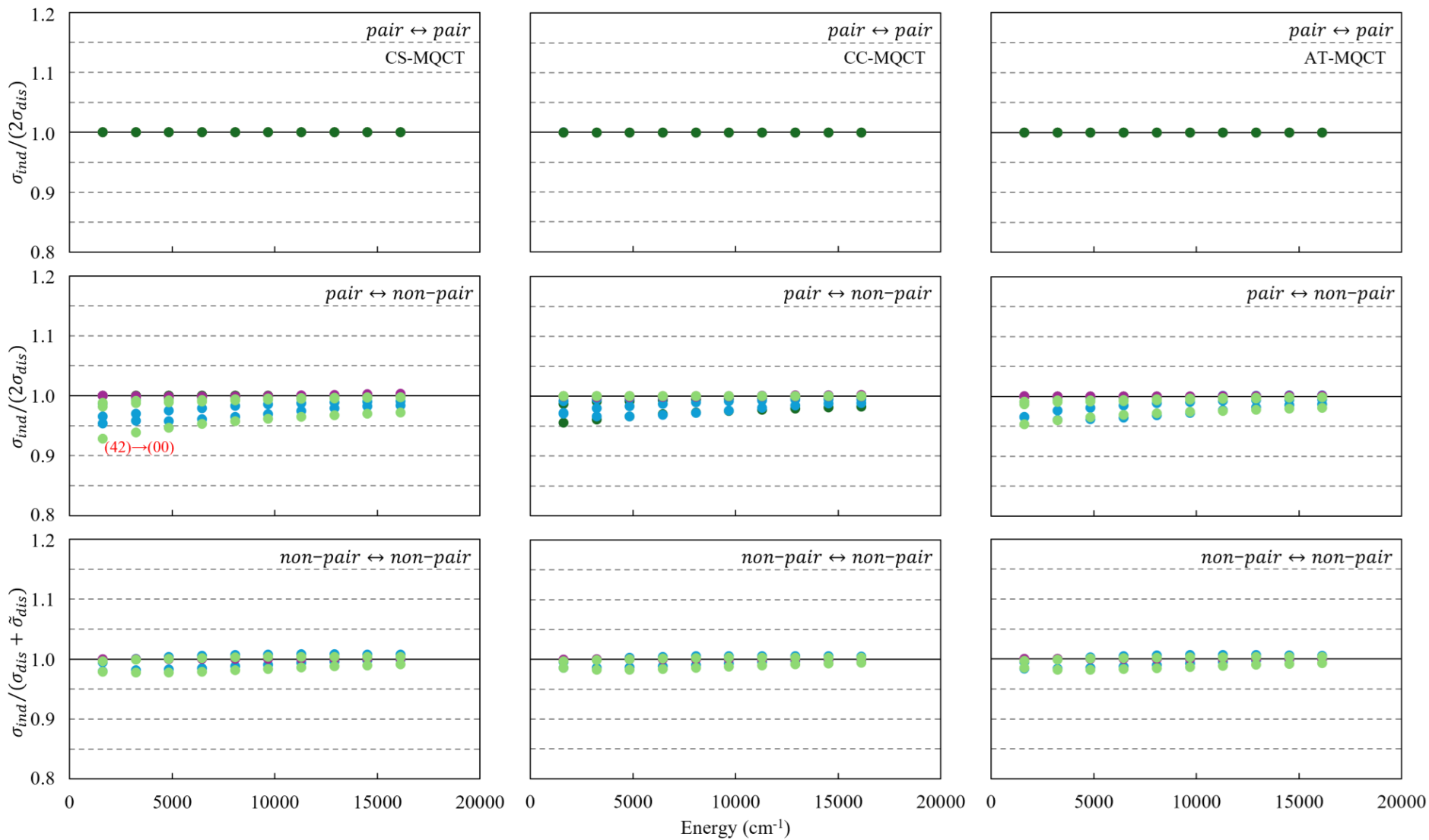

**Figure S1:** Ratios of inelastic cross sections for $H_2 + H_2$ obtained by treating collision partners either as indistinguishable or distinguishable in MQCT calculations at different collision energies. Top frames display the data for pair ↔ pair transition, middle frames for pair ↔ non-pair and lower frames for non-pair ↔ non-pair transitions. Results are obtained using three different versions of MQCT methods: CS-MQCT (left frames), CC-MQCT (middle frames) and AT-MQCT (right frames). Transitions are grouped by their initial states, and each initial state is assigned a certain color: blue for (00), orange for (22), dark green for (44), light blue for (20), violet for (40), light green for (42). The full list of presented transitions is given in Table S3.

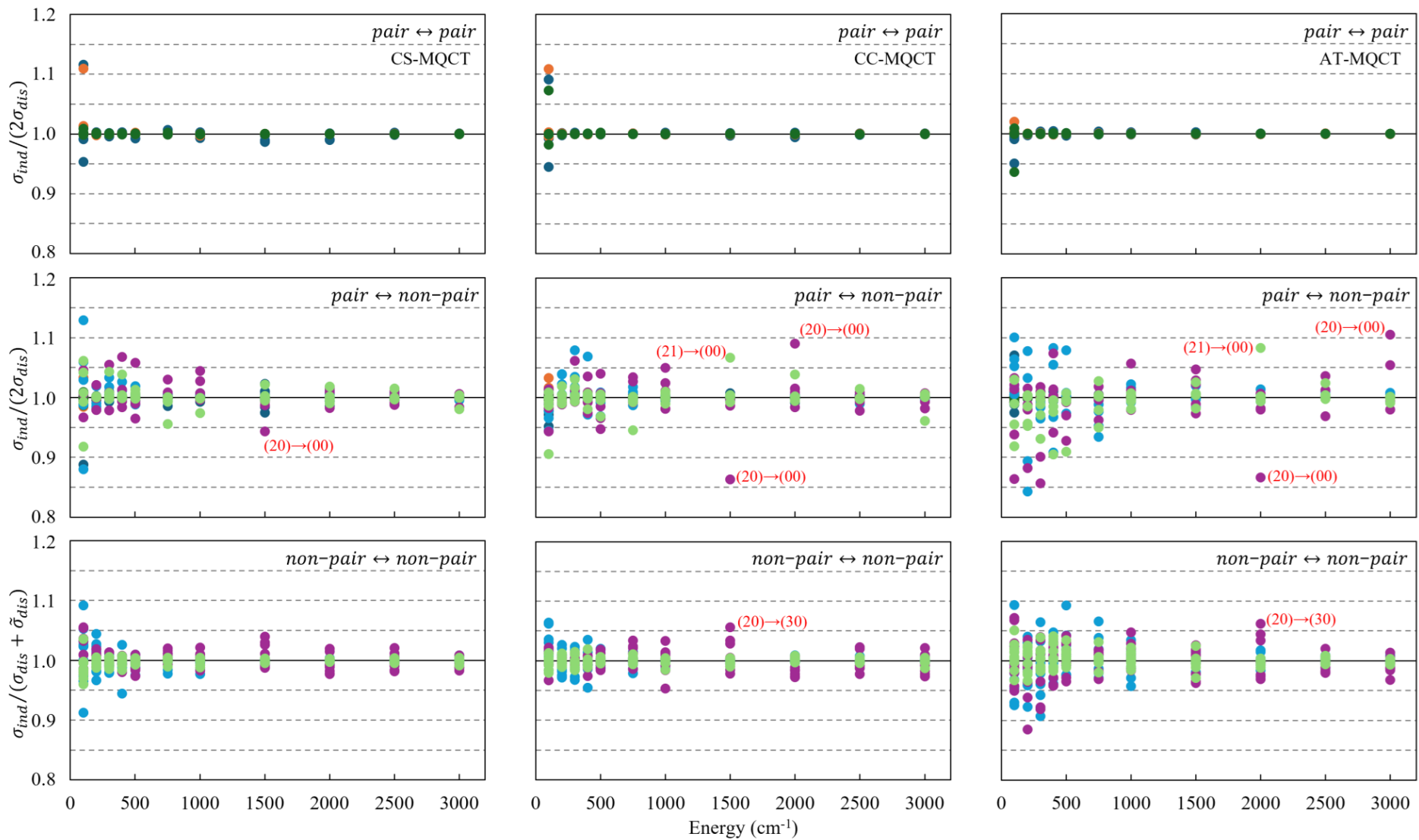

**Figure S2:** Same as Fig. S1, but for CO + CO system. Transitions are grouped by their initial states, and each initial state is assigned a certain color: blue for (00), orange for (11), dark green for (22), light blue for (10), violet for (20), light green for (22). The full list of presented transitions is given in Table S3.

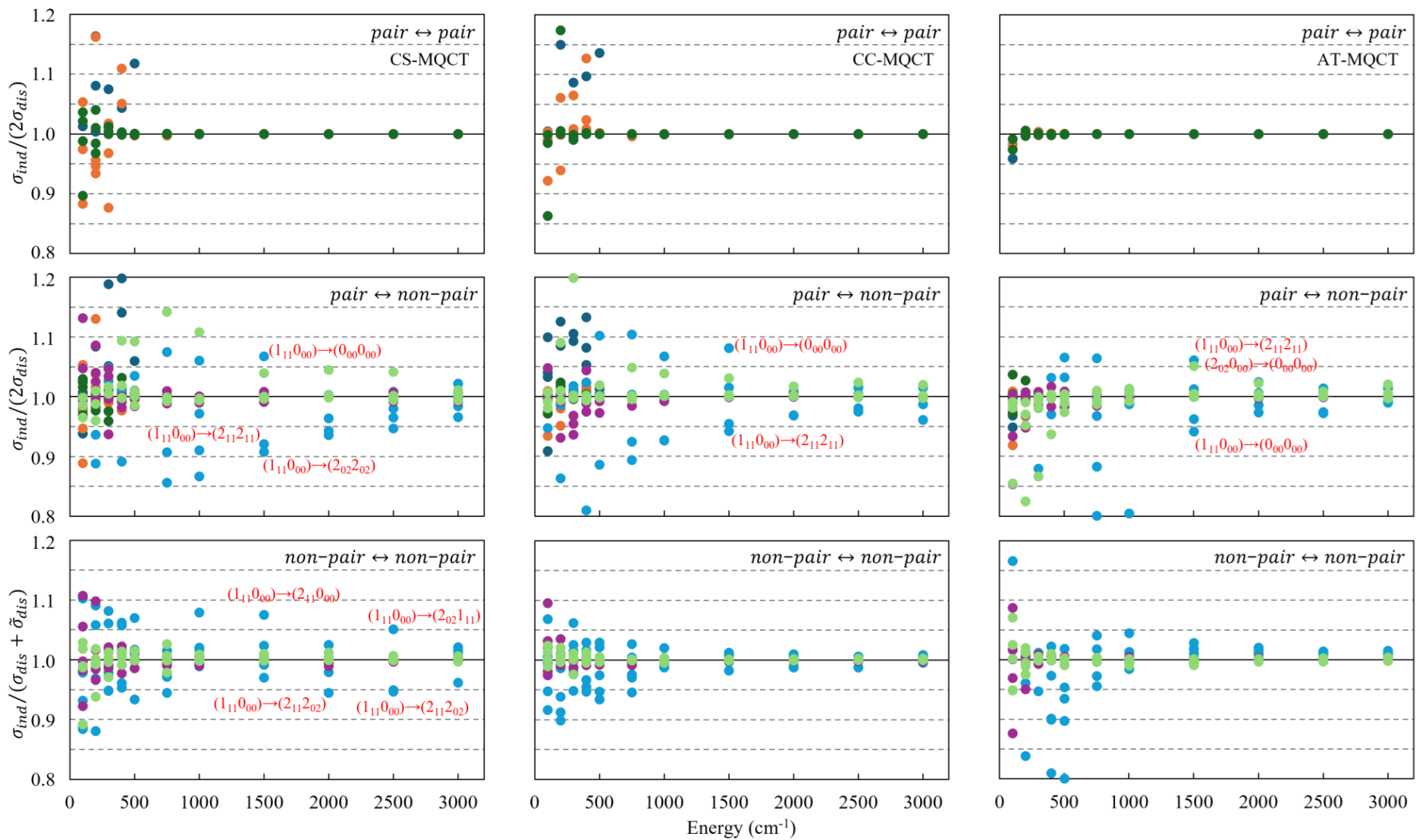

**Figure S3:** Same as in Fig. S1, but for *p*-$H_2O$ + *p*-$H_2O$ system. Transitions are grouped by their initial states, and each initial state is assigned a certain color: blue for $(0_{00}0_{00})$, orange for $(1_{11}1_{11})$, dark green for $(2_{02}2_{02})$, light blue for $(1_{11}0_{00})$, violet for $(2_{02}0_{00})$, light green for $(2_{02}1_{11})$. The full list of presented transitions is given in Table S3.

37